\documentclass[preprint,showpacs,preprintnumbers,amsmath,amssymb,superscriptaddress,floatfix]{revtex4}
\usepackage{graphicx}
\begin{document}

\title{XRD and micro-Raman study of structural transformations in\\
(B$_2$O$_3$)$_{1-x}$(H$_2$O)$_x$ glasses and liquids}
\author{Ralf Br\"uning}
\email[]{rbruening@mta.ca, Tel.: (506) 364-2587, Fax.: (506) 364-2583}
\author{Justine B. Galbraith}
\author{Katherine E. Braedley}
\author{Jonathan Johnstone}

\affiliation{Physics Department, Mount Allison University, Sackville, New Brunswick,
Canada E4L 1E6}

\author{Jacques Robichaud}
\author{Subramanian Balaji}
\author{Yahia Djaoued}
\affiliation{Laboratoire de Micrco-Spectrocopie Raman et FTIR,
Universit\'e de Moncton -- Campus de Shippagan,\\
218, boul.\ J.-D. Gauthier,
Shippagan, New Brunswick E8S 1P6, Canada }


\date{\today}

\begin{abstract}

Liquid water and vitreous B$_2$O$_3$ are the endpoints of a continuous range of random networks in which hydrogen bonds gradually replace covalent bonds.  Previous work has shown that glasses can be obtained by quenching in the range $x \le$ 0.50. We report the wide-angle x-ray scattering by the liquid phase in the composition range from $x$ = 0.38 to $x$ = 1.00 (pure water) at temperatures just above the liquidus. The first sharp diffraction peak (FSDP) remains at an approximately constant position in the range from
$0 \le x \le 0.8$. Beyond this range, the position of the FSDP shifts  linearly to higher angles.   The relative concentration of the molecular species in the glasses and melts were measured with micro-Raman spectroscopy.
Small molecular species are found for glasses and liquids with $x > 0.36$, determining the critical point at which the sample ceases to be a single macromolecule.  Molecular water is present in liquids
with $x > 0.62$.
\end{abstract}

\maketitle

\section{\label{Introduction} Introduction}
  
Boron oxide and water react spontaneously according to $(1-x){\rm B_2O_3} + x{\rm H_2O} \to {\rm B}_{{\rm 2-2}x}{\rm O}_{{\rm 3-2}x}{\rm H}_{{\rm 2}x}.$
Pure B$_2$O$_3$ invariably forms a glass upon cooling. With increasing water content, hydrogen bonds gradually replace the continuous random network of covalent B--O bonds of vitreous B$_2$O$_3$, and eventually the network breaks up into various borate species.  
Krazek et al.\ measured the liquidus temperature and vapor pressure, and they identified the crystalline phases \cite{kra}.  Glasses with $x \le 0.39$ have been examined by x-ray scattering \cite{mil,mel}, NMR \cite{sil,mil1}, IR and Raman spectroscopy \cite{par}.  In these glasses molecular water dissociates into hydroxyl groups \cite{sil,par}, and the strength of the hydrogen bonds increases with water content \cite{par}.
The density as a function of composition decreases first gradually for $x<0.25$, then more rapidly for higher $x$ \cite{kod,mil1}.  Glasses can be obtained in the range from $x$=0.00 to $x$=0.50; outside this range B(OH)$_3$ crystals appear even upon quenching \cite{bru4}.   At room temperature B$_2$O$_3$ glass absorbs water up to $x=0.41$.  Beyond this concentration the water-saturated glass is in metastable equilibrium with crystalline B(OH)$_3$.  Crystals of  $\alpha$-HBO$_2$, consisting of B$_3$O$_3$(OH)$_3$ molecules, appear only upon heating past 70$^{\circ}$C \cite{mcc}.

Relatively little is known about the water-borate system above the glass forming range. Schmidt et al.\ studied aqueous samples with concentrations 0.959$ \le x \le $0.996 at temperatures between 25$^{\circ}$C and 500$^{\circ}$C and pressures from 0.1 MPa to $\sim$2 GPa \cite{sch}. In these water-rich solutions, B(OH)$_3$ is the main molecular species present at all temperatures and pressures. Lesser amounts of a tetrahedrally coordinated boron species, similar to $\beta-$HBO$_2$, are present in compressed liquids above 100$^{\circ}$C \cite{sch,tho}.

Here we report the results of wide-angle x-ray scattering measurements in the liquid phase for 0.38 $\le x \le$ 1.00 that complement previously published results for glasses with $x\le0.501$ \cite{bru4}.  Raman spectroscopy measurements were carried out in the entire composition range.  The data provide information on the distribution of different molecular species, and they indicate the degree of borate polymerization as a function of composition. Calorimetric measurements of the glass transition indicate a progression from strong to fragile behavior as $x$ approaches 0.5, with a rapid increase of the excess specific heat of the supercooled liquid for glasses with $x>0.3$ \cite{bru4}.  Fragile systems are typically ionic or molecular liquids \cite{ang1,boh}, and the results reported here indicate that small molecules are indeed present in water-rich glasses.

\section{\label{experimental} Experimental Methods}

 A series of (B$_2$O$_3$)$_{1-x}$(H$_2$O)$_x$ mixtures ($0.381 
\le x \le 1.000$) were prepared from boron oxide powder (Alfa Aesar, 99.98\% 
metals basis) and deionized water ($\sim$20 ppm impurities). The initial water content of the boron oxide was found to be $x=0.0784$, as determined by weight loss upon heating to $900^{\circ}$C. Samples were prepared by partially filling capillaries of vitreous silica with weighed amounts of water and boron oxide. The capillaries (0.7 mm diameter, 0.010 mm wall thickness, W.\ M\"uller, Sch\"onwalde) can withstand the vapor pressure of the samples at the liquidus temperatures of up to 0.7 MPa. The open ends of the capillaries were sealed in a flame, and the capillaries were weighed before and after sealing to ensure that no significant amount of water had been lost.
The sealed capillaries were placed horizontally in a furnace and heated overnight to melt and mix the material.
They are partially filled, with volume fractions ranging from 45 to 86 $\%$, and compositions have an uncertainty of $\pm$0.005. 

The x-ray scattering of the heated mixtures at their liquidus temperature was measured with a custom-built diffractometer equipped with graphite monochromator and analyzer crystals and using CuK$_{\alpha}$ radiation. The experimental method employed for measuring the glasses, by scattering of a flat sample surface in reflection, has been described previously \cite{bru4}.  For the present measurements on melts we have constructed an x-ray mirror furnace with a halogen lamp at one focal line of an elliptical mirror.  The samples were placed along the second focal line inside an evacuated chamber.  A Kapton window allows the x-rays to pass, while the thermal radiation enters through a heat-resistant glass window.  Samples were heated until completely melted by visual inspection.  In the line of the x-ray beam, capillaries contained both air pockets and liquid sample.  Temperatures could not be determined reliably in this geometry, and we assume that the sample temperature equals the known liquidus temperature for each composition (between 60 and 250$^{\circ}$C, with vapor pressures between 2 to 710 kPa \cite{kra}).  Empty capillary signals, contributing up to 60\% of the overall signal, were measured at the relevant furnace power settings.  These empty capillary signals were reduced to allow for the absorption by the sample and then subtracted to obtain the net sample scattering.
An example of the empty capillary subtraction is shown in Fig.\ 1.
The data, measured as a function of scattering angle  $2 \theta$, are shown as a function of the modulus of the scattering vector $q = 4 \pi \lambda^{-1} \sin{\theta}$, where  $\lambda = 0.15418\, {\rm nm}$ is the wavelength.

\begin{figure}[ht]	           
\includegraphics[scale=1.0]{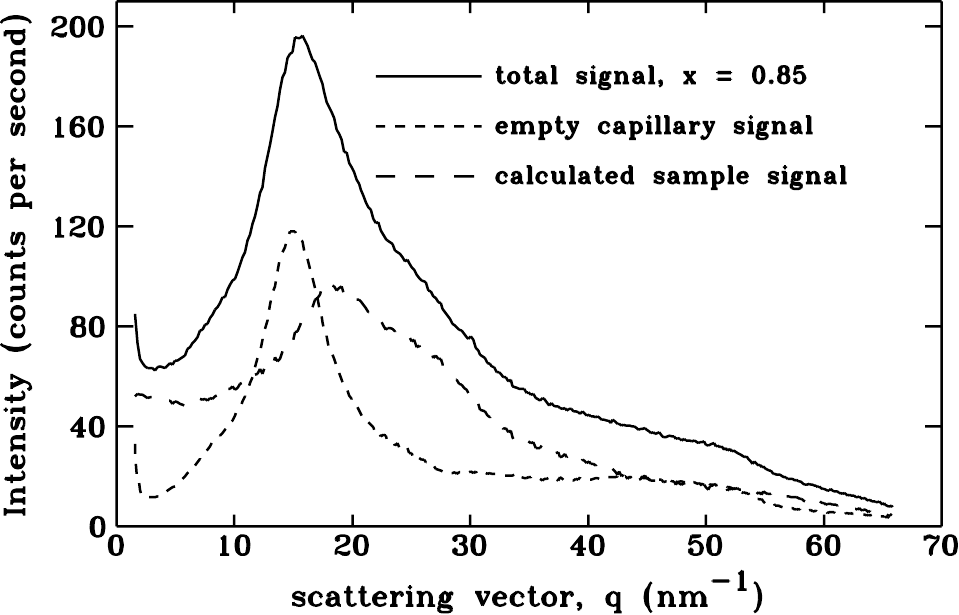} 
\caption{\label{pdiffer1}
Observed scattering for full and empty capillaries at T $\gtrsim$ 430 K. The approximate net sample scattering is obtained by subtracting 90$\%$ of the empty capillary signal.
}
\end{figure}

In the further analysis of the x-ray scattering data, a polarization factor accounts for the partial polarization of the x-ray beam, and the round geometry of the capillaries necessitates an absorption correction \cite{bra}.  The result is the observed scattering intensity, $I_{obs}(q)$. The x-ray count rate is converted to electron units by dividing with a factor, $N$, and the tabulated incoherent scattering, $I_{inc}(q)$, is subtracted \cite{war} in order to obtain the coherent scattering,
$I(q) = {I_{obs}(q)/N-I_{inc}(q)}$.  The composition average, $\langle f^2(q) \rangle$, of the atomic scattering factors $f_{\rm B}(q), f_{\rm O}(q)$, and $f_{\rm H}(q)$ was calculated  \cite{cro}. The value at $q=0$ was used to normalize the coherent scattering according to $I'(q) = I(q) / \langle f^2(0)\rangle$ in order to facilitate the visual comparison of data at different compositions.

Micro-Raman analysis was performed with a Jobin-Yvon Labram HR microanalytical spectrometer equipped with a motorized x-y stage. The spectra were generated with 17 mW, 632.8 nm He-Ne laser excitation and were dispersed with a 1800 gr/mm grating across the 0.8 m length of the spectrograph.  A 10$\times$  objective was used, and the laser power was 5.3 mW on the sample surface.  The spectral resolution is estimated to be better than 0.5 cm$^{-1}$ for a slit width of 150 $\mu$m  and a confocal hole of 300 $\mu$m.  The capillaries were heated inside a furnace mounted on the microscope stage.  The furnace consisted of three pieces of concentrically aligned stainless steel syringe tubing, with Teflon-coated heating wire between the two innermost steel tubes.  Optical access was provided by aligned holes with a size of about 1 mm, and a thermocouple assembly abutted one end of the capillaries.
				
\section{\label{resultsx} X-ray Scattering Results}

\begin{figure}[ht]	           
\includegraphics[scale=1.05]{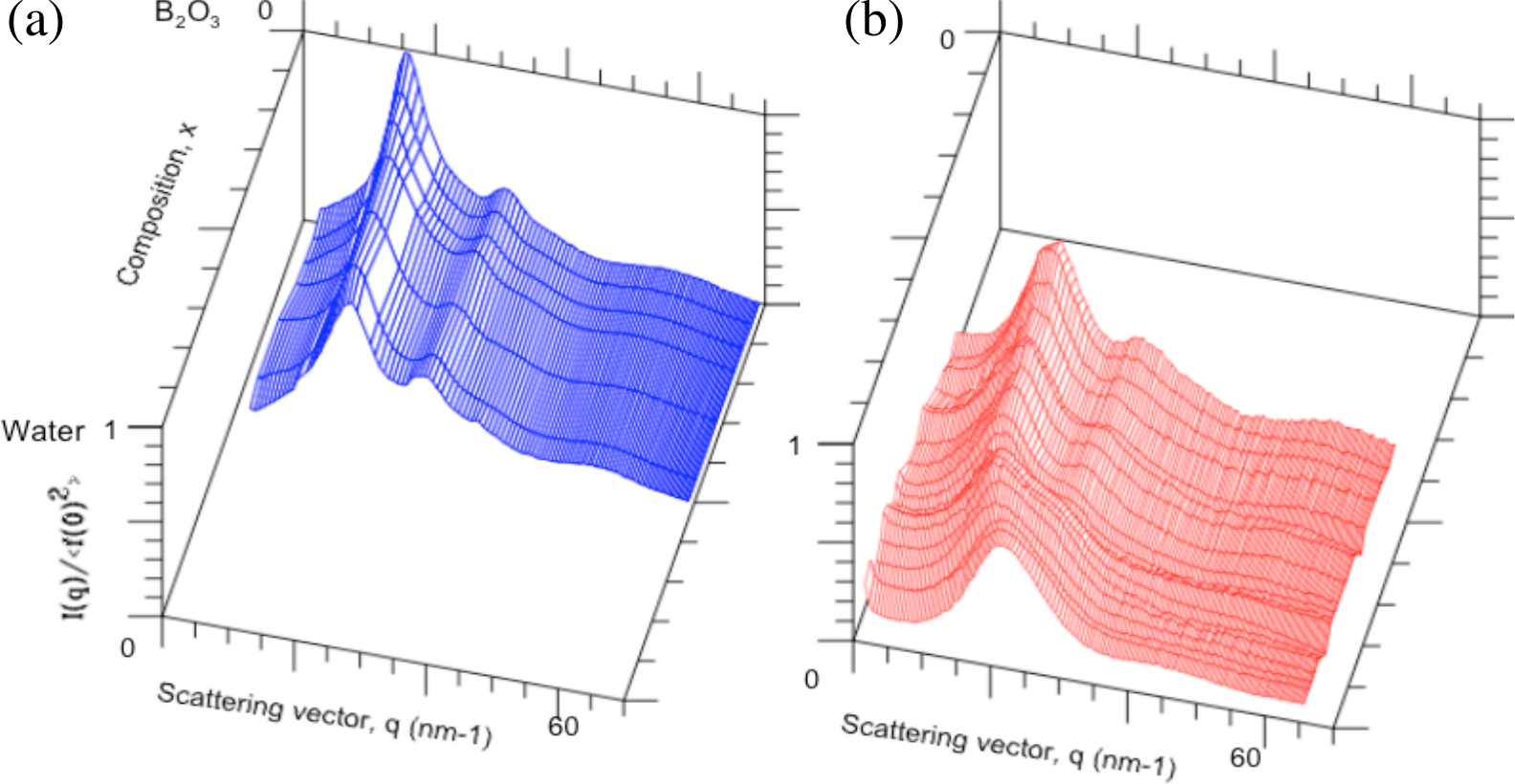} 
\caption{\label{melt3d}
Coherent x-ray scattering by (a) (B$_2$O$_3$)$_{1-x}$(H$_2$O)$_x$ glasses for $x \leq 0.501$ at room temperature and (b) liquids for    
$x \geq 0.381$.  With the exception of pure water and     
$x=0.964$, the liquid mixtures were measured at their respective liquidus temperatures.  Intensities are normalized to $\langle f(q=0)^2 \rangle$.}
\end{figure}

\begin{figure}[ht]	           
\includegraphics[scale=1.0]{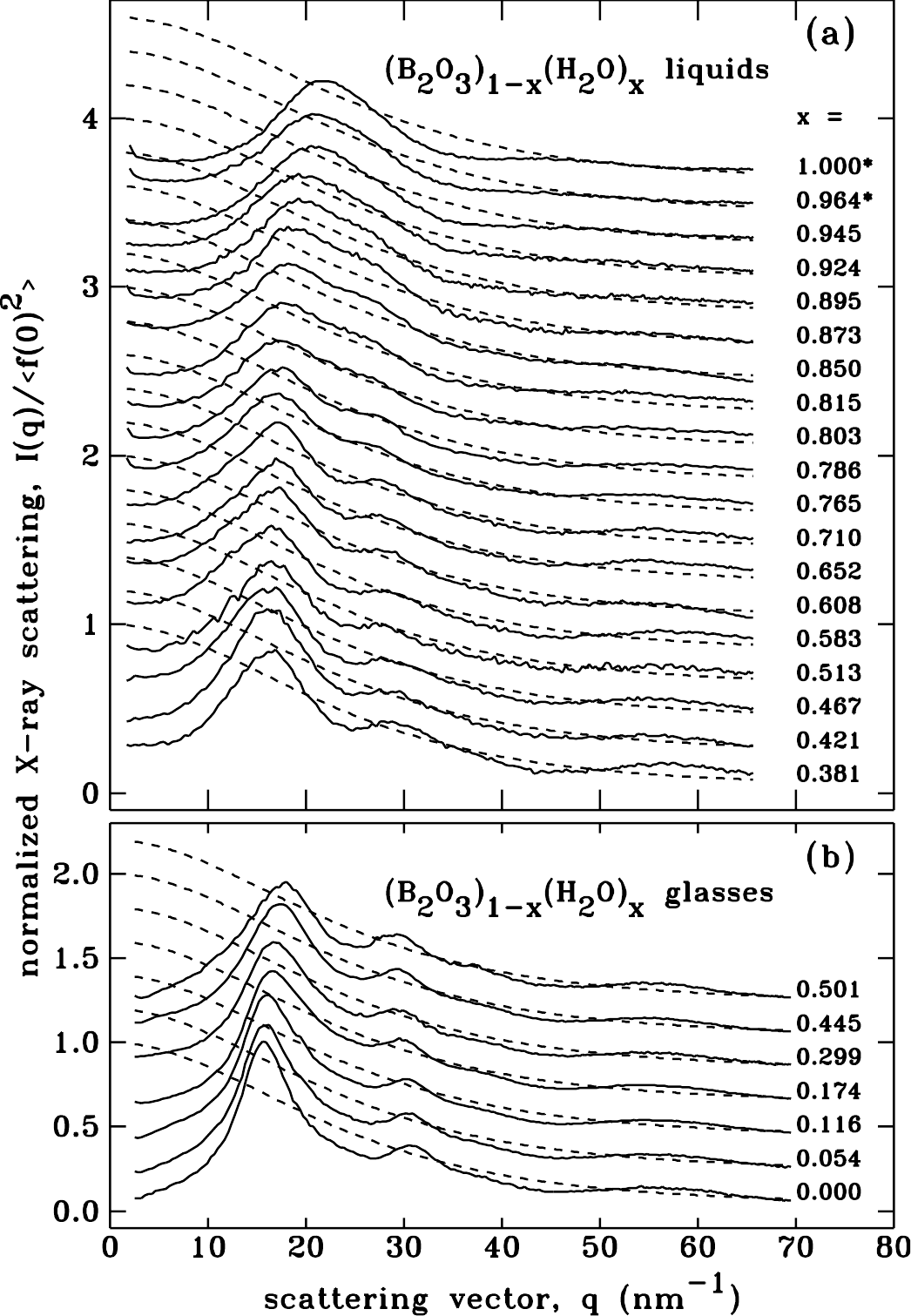} 
\caption{\label{all}
Coherent x-ray scattering by (a) (B$_2$O$_3$)$_{1-x}$(H$_2$O)$_x$ glasses at room temperature and (b) liquids.  Dashed lines represent the ratio $\langle f(q)^2 \rangle / \langle f(q)^2 \rangle$.
With the exception of pure water and     
$x=0.964$ (*), the liquids were measured at their respective liquidus temperatures.  Successive curves are shifted by intervals of 0.2.
}
\end{figure}

 Figure \ref{melt3d} shows the coherent component of the x-ray scattering as a surface plot for the full composition range from B$_2$O$_3$ to H$_2$O.  Here the normalized scattering intensity, $I'$, is shown as a function of $q$ and composition, $x$.  The data for glasses at room temperature in figure \ref{melt3d}(a) are redrawn from reference \cite{bru4}.  They cover the glass forming range, which extends up to near $x \le 0.501$.  Figure  \ref{melt3d}(b) shows the scattering of water-rich mixtures ($0.381 \le x \le 0.945$) heated to their liquidus point.  However, for pure water and $x=0.964$ patterns measured above the $T_l$ at about 423 K were plotted,
because otherwise the effect of the steep decrease of $T_l$ in the vicinity of $x=1$ would be somewhat misleading.  (The temperature dependence of the scattering of water is discussed below.)  The same data are shown as conventional plots in figures \ref{all}(a) and (b).  The results for the endpoints $x=0$ and $x=1$ agree with the results obtained by other groups \cite{cha,hur}.  Above $q = 50 \, {\rm nm}^{-1}$ the precision of the data for the liquids is limited by the large contribution of the empty capillary scattering.

\begin{figure}[ht]	           
\includegraphics[scale=0.6]{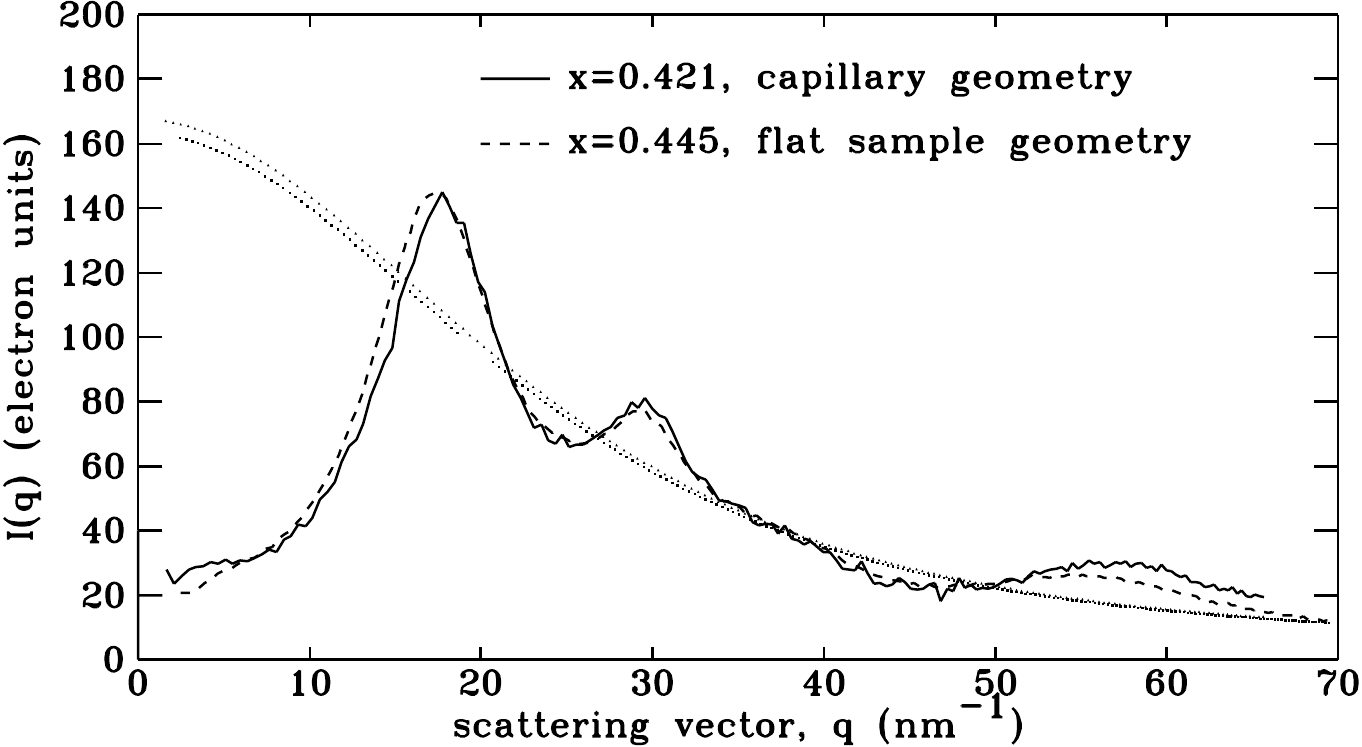} 
\caption{\label{pfr}
Scattering by glasses with similar compositions at room temperature. The $x=0.421$ glass was enclosed in a capillary, while the $x=0.445$ sample was a flat piece of glass in vacuum. Dotted lines show  $\langle f^2(q) \rangle$ for the two compositions. 
}
\end{figure}

Glasses with $x$=0.381, 0.421 and 0.467 were prepared in capillaries by quenching in air from above the liquidus temperature to ambient temperature.  Figure \ref{pfr} compares the scattering of two samples with similar composition with different geometries.  We conclude that consistent results are obtained with these two scattering geometries. 

\begin{figure}[ht]	           
\includegraphics[scale=0.6]{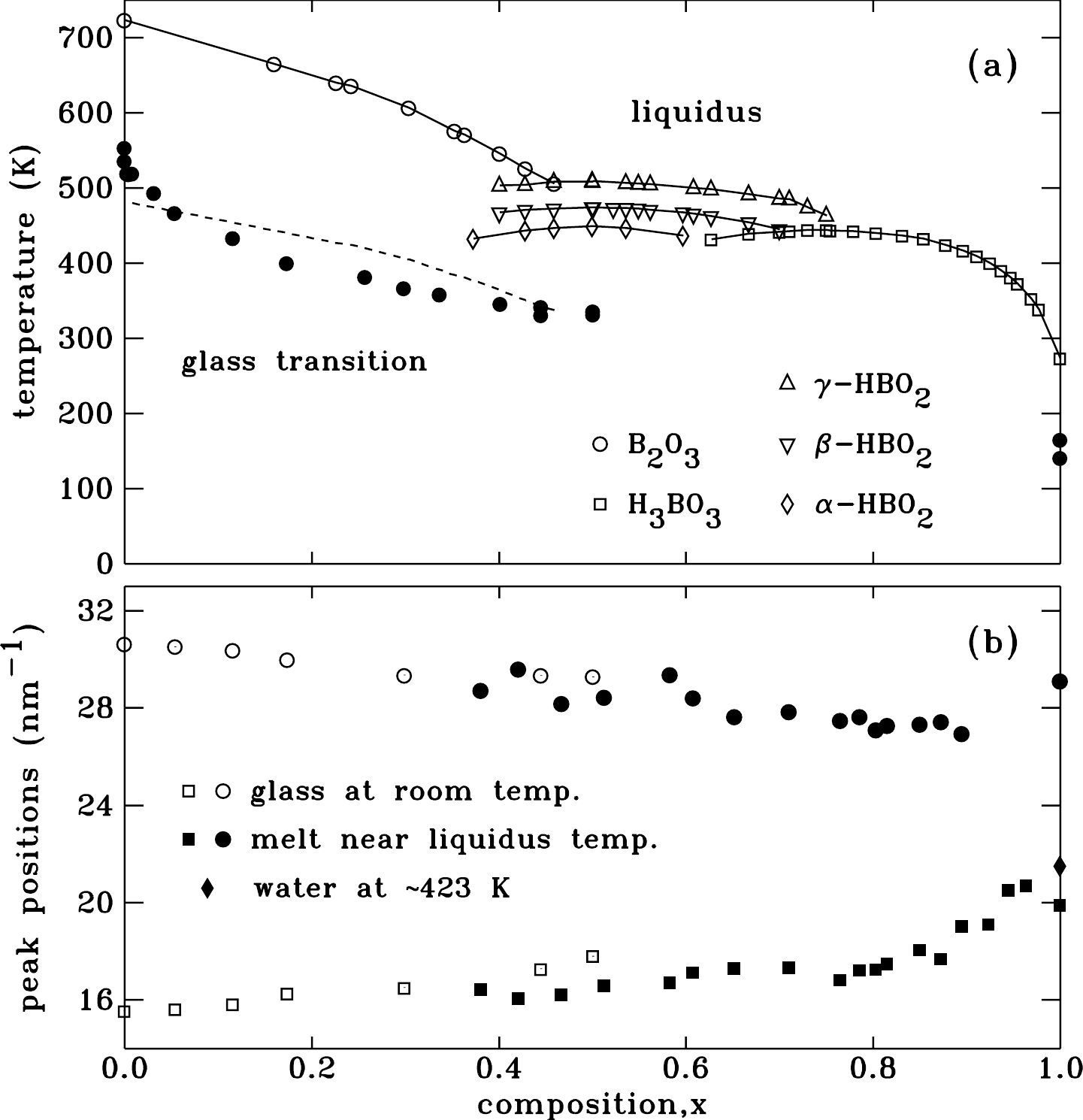} 
\caption{\label{liquidus}
(a) Liquidus and glass transition temperatures for the (B$_2$O$_3$)$_{1-x}$(H$_2$O)$_x$
system redrawn from references \cite{kra,bru4,vel}.  The dashed line at
2/3 of $T_l$ coincides approximately with $T_g$. (b) Positions of the maxima of the x-ray scattering as a function of $x$.}
\end{figure}

\begin{figure}[ht]	           
\includegraphics[scale=0.6]{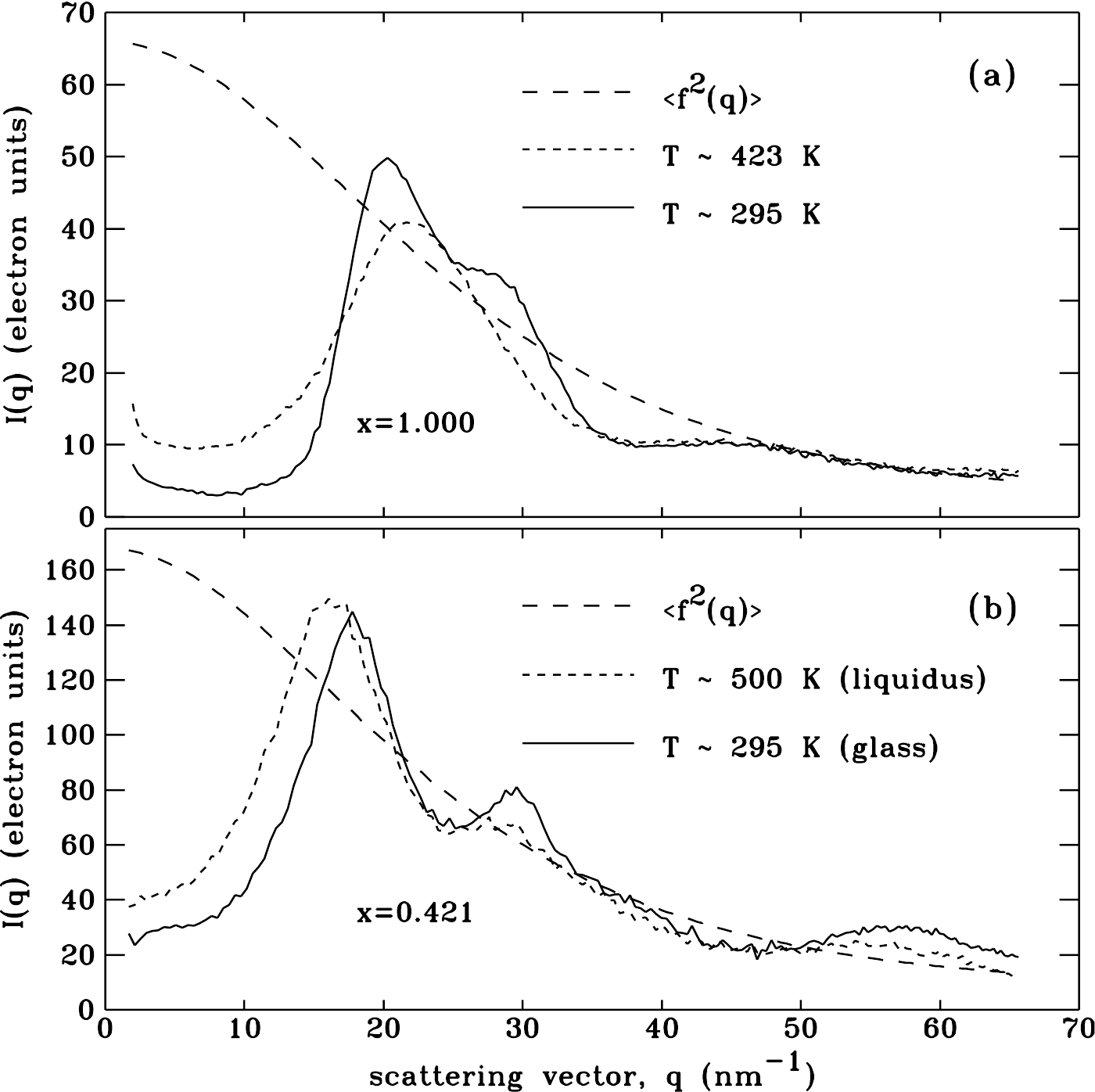} 
\caption{\label{421water}
(a) Coherent component of the x-ray scattering of water at two temperatures.  (b) Same for a mixture with
$x=0.421$ measuread as a glass at room temperature and as a melt just above the liquidus temperature.
All data were obtained with samples in sealed capillaries.
}
\end{figure}

Figure \ref{liquidus}(a) shows the liquidus and glass transition temperatures of (B$_2$O$_3$)$_{1-x}$(H$_2$O)$_x$ mixtures as a function of composition.  We find approximate agreement with the empirical rule that the glass transition temperature, $T_g $, is two thirds of the melting point \cite{kau}. Pure B$_2$O$_3$ exhibits three broad diffraction peaks in the $q$-range considered here.  The intensity of the first sharp diffraction peak (FSDP), centered around 17 nm$^{-1}$,  decreases with increasing $x$.  This decrease is steepest in the range from $x=0$ to $x=0.3$ [figure \ref{all}(b)].  We note that the decrease of the FSDP coincides with the drop in $T_g$ as water is added to B$_2$O$_3$ (figure \ref{liquidus}).  The height of the normalized
FSDP for the liquids remains nearly constant in the
range from $x=0.381$ to $0.710$.  At higher $x$ the FSDP merges with the second broad peak.  A single broad FSDP is characteristic of pure {\it hot} water \cite{hur}, while water at room temperature exhibits as FSDP with a clearly defined shoulder [figure \ref{421water}(a)].

Figure \ref{liquidus}(b) shows the $q$ values where the sample scattering, relative to the curve $\langle f(q)^2 \rangle$, assumes a maximum.  The position of the FSDP remains close to 16 nm$^{-1}$ for $x \le 0.8$.  For water-rich compositions with
$x \ge 0.8$ the FSDP shifts linearly from $q = 17 \, {\rm nm}^{-1}$ to $21 \, {\rm nm}^{-1}$.
The position of the second peak decreases linearly with $x$ from $30.5 \, {\rm nm}^{-1}$ to $27 \, {\rm nm}^{-1}$ for both vitreous and liquid samples.  Beyond about $x=0.9$ the first and second peak overlap to a degree that they can no longer be distinguished.  As shown in figure \ref{421water}(a), room-temperature water displays a second peak in this range.  The position of the second room temperature water peak is clearly not consistent with the extrapolated peak positions of the borate melts.

The scattering signals of B$_2$O$_3$-rich and water-rich compositions respond differently to 
temperature changes.  The scattering of pure liquid water changes dramatically upon heating, with the first and second diffraction peak merging into a single peak at 423 K (figure \ref{421water}(a)).  At 350 K one still observes 
shoulder at the right side of the first sharp diffraction peak of water \cite{hur1}.  As expected, the scattering at small angles increases upon heating due to the increasing density fluctuations.
Figure \ref{421water}(b) shows data for a sample with $x= 0.421$ measured both in its vitreous state at room temperature, about 35 K below
$T_g =330$ K \cite{bru4}, and its liquid state near 500 K.  For the liquid state the 
amplitudes of the second and third peak are lower, consistent with a loss of short-range order upon heating.  Contrarily, the area of FSDP grows, indicating that 
the liquid can have a degree of long-range order that exceeds that of the glass with the same 
composition.  The origin of this may be that the structure is less frustrated when some of the bonds are broken at elevated temperature.  Small angle scattering increases again with increasing temperature.

\section{\label{resultsr} Raman Spectroscopy Results}

\begin{figure}[ht]	           
\includegraphics[scale=0.8]{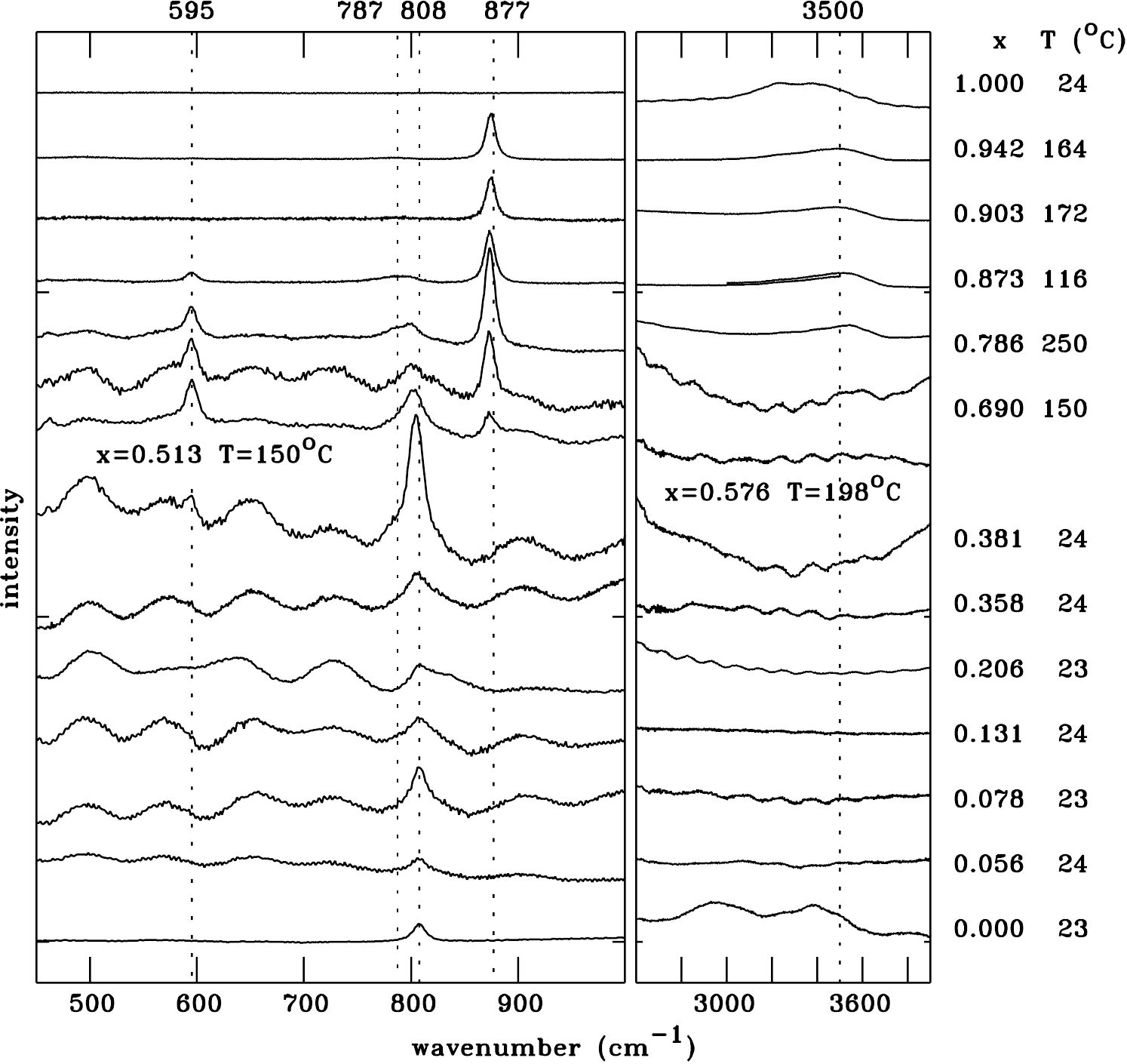} 
\caption{\label{ram1}
Raman spectra of (B$_2$O$_3$)$_{1-x}$(H$_2$O)$_x$ glasses ($x \le 0.381$) and liquids ($x \ge 0.513$).
The spectra are shifted vertically for clarity.  Dashed lines mark the nominal positions of the peaks used in the analysis.
}
\end{figure}

\begin{figure}[ht]	           
\includegraphics[scale=1.0]{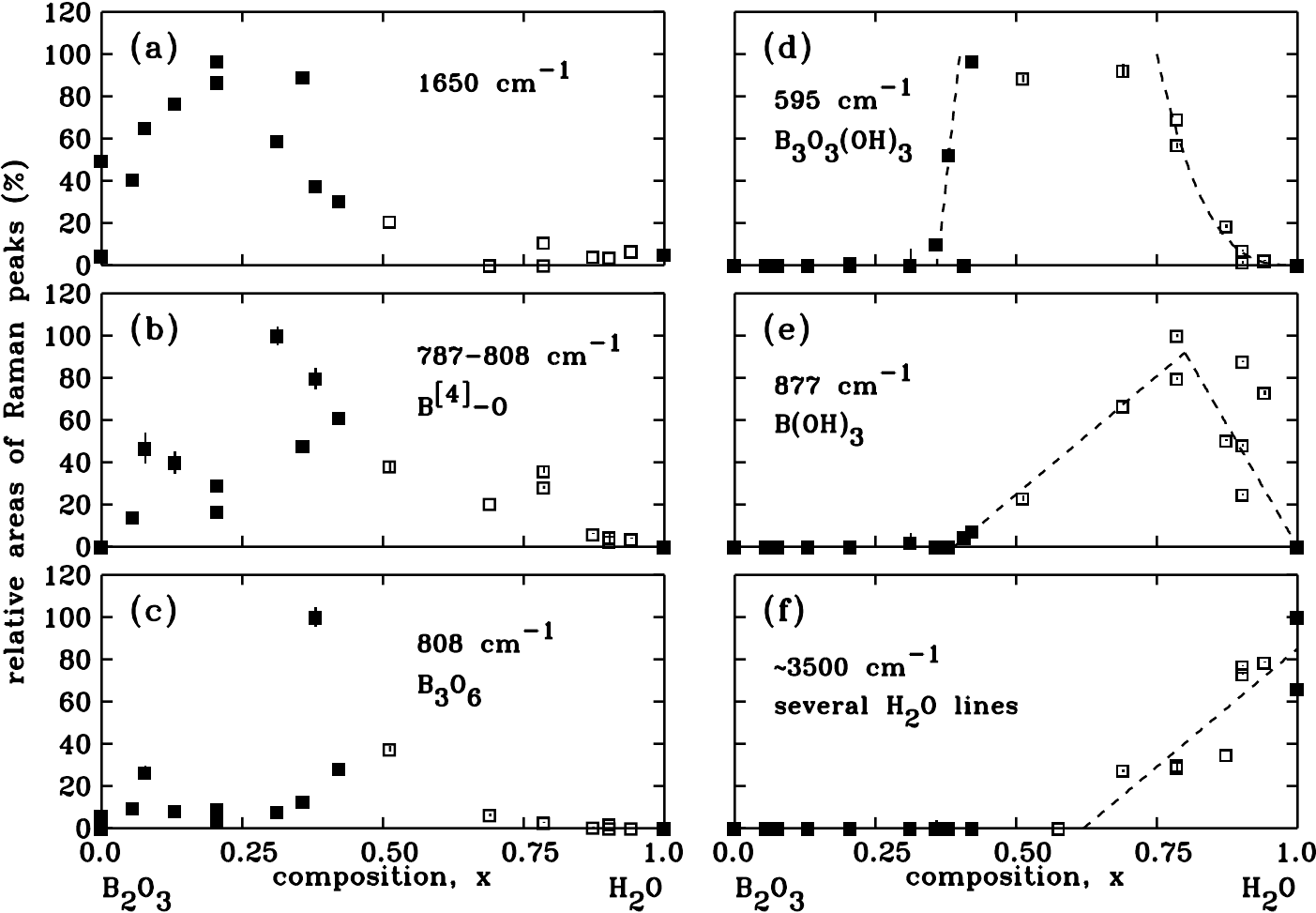}
\caption{\label{ram2}
Relative areas of selected Raman peaks, normalized to the highest observed value for each line. The peak center and the molecular species assigned to each peak are indicated in panels (a)-(f). Room temperature and liquidus temperature data are shown by full and open symbols, respectively.
}
\end{figure}

Raman spectroscopy allows one to identify molecular species through their distinct vibrational resonances \cite{sch}.  Therefore the degree of polymerization can be monitored as a function of composition.
Table \ref{freq} lists the Raman frequencies with their assigned molecular species and vibrational modes.
Figure \ref{ram1} shows selected Raman spectra obtained for glasses with $x \le 0.422$, as well as the spectra with $x \ge 0.513$ at the liquidus temperature.  In the following
we consider the intensity of six Raman lines that can be identified most clearly.  They are marked by dashed lines in figure \ref{ram1} and the composition dependence of their areas is
shown in figure \ref{ram2}.  The areas were determined by least squares fitting to Gaussian peaks.

\begin{table}[t]
\caption{Known Raman frequencies, $\nu$, and modes of various molecular species.}
\label{freq}
\begin{center}
\begin{tabular}{c|l}
$\nu$ (cm$^{-1}$) & Molecular species and mode if known\\ \hline \hline

187, 211 & Lattice modes, librations of B(OH)$_3$ molecules \cite{kuz} \\
380-500 & Symmetric stretching of B$^{[4]}$-O in hydrated polyborate anions \cite{jun} \\
500 & Symmetric B-O vibration of B$^{[3]}$-O in planar B(OH)$_3$ \cite{tho, jan, sch, jun} \\
550 & Deformation of the O-B-O bond of B$^{[3]}$-O in B(OH)$_3$ \cite{kuz, jan, par} \\
568 & [B$_4$O$_5$(OH)$_4$]$^-$ ion \cite{sch} \\ 
595 & B$_3$O$_3$(OH)$_3$ metaboric acid molecule \cite{par, gal} \\
613 & [B$_3$O$_3$(OH)$_4$]$^-$ ion \cite{sch} \\
650-665 & Out-of-plane stretching of the covalent bonds of B$^{[3]}$-O in B$_2$O$_3$ \cite{sim, jan} \\
744 & [B(OH)$_4$]$^-$ orthoborate ion \cite{tho, sch} \\
767 & [B$_5$O$_6$(OH)$_4$]$^-$ pentaborate ion \cite{tho, sch} \\
780-790 & Symmetric stretching mode of B$^{[4]}$-O \cite{sch, jun,jan} \\
808 & Symmetric stretching mode of boroxol rings, B$_3$O$_6$ \cite{gal,sim,hua} \\
877 & Symmetric stretching vibration of B$^{[3]}$-O in planar B(OH)$_3$ \cite{tho, kuz, sch, jun, jan}  \\
1000-1150 & Asymmetric stretching of B$^{[4]}$-O in hydrated polyborate anions \cite{jun} \\
1167 & In-plane B-O-H angle deformation mode in B(OH)$_3$ \cite{kuz, jun, par} \\
1600 & Deformation of H$_2$O molecules \cite{rav} \\
3000-3600 & O-H stretching vibration bond \cite{tho, jun, rav, kuz} \\
3168 & \ \ \ \ \ \ \ \ \ Symmetric O-H stretching in B(OH)$_3$ \cite{kuz, tho} \\
3245 & \ \ \ \ \ \ \ \ \ Antisymmetric O-H stretching in B(OH)$_3$ \cite{kuz, tho} \\
3500 & \ \ \ \ \ \ \ \ \ O-H stretching in water molecules \cite{tho,rav}
\end{tabular}
\end{center}
\end{table}

The asymmetric group of lines near 3500 cm$^{-1}$ can be used to identify molecular water \cite{tho}.
The shape of this broad feature was fitted to a broad and a narrow Gaussian peak, and we consider the sum of the areas of these two Gaussian profiles [figure \ref{ram2}(f)].
According to this consideration, the limiting composition beyond which we observe free molecular water is between     
$x=0.576$ and 0.690.   Within the experimental uncertainty the area of the Raman peaks at 3500 cm$^{-1}$ increases linearly with $x$, and a linear extrapolation of the peak areas gives $x \approx 0.62$ as the limit for the presence of H$_2$O molecules.  We conclude that in the liquid state molecular water persists below the stoichiometric composition of B(OH)$_3$ at $x=3/4$. 

The B(OH)$_3$ molecule can be considered as the monomer that reacts to form an extended 
B$_{2-2x}$O$_{3-2x}$H$_{2x}$ network through a condensation reaction.
The sharp Raman line at 877 cm$^{-1}$ is the strongest feature in the water-rich liquids (figure \ref{ram1}), and it is associated with the symmetric stretching frequency of the BÐO bond of the B(OH)$_3$ molecule \cite{sch}.  The intensity of this line assumes a maximum at a composition near $x \approx 0.80$, i.e.\ slightly above $x=3/4$ [figure \ref{ram2}(e)].  At and below $x = 0.38$ (room temperature spectra) 
this Raman line can no longer be detected, and we conclude that all monomers have polymerized into larger units.  Conversely, B(OH)$_3$ molecules persist in water-rich glasses with $x>0.38$.

A sharp Raman line at 595 cm$^{-1}$ has been associated with B$_3$O$_3$(OH)$_3$ rings \cite{par,gal}.  These rings are the structural unit of the low-temperature form of metaboric acid, $\alpha$-HBO$_2$, with stoichiometry $x=1/2$.  One can see in figures \ref{ram1} and \ref{ram2}(d) that this line is indeed prominent in liquids and glasses with compositions in the vicinity of
$x=0.5$.  The B$_3$O$_3$(OH)$_3$ molecule can be considered at a borate trimer, and we expect
that at low borate concentrations the occurrence of trimers should be proportional to the cube of the monomer concentration.  The dashed line in figure \ref{ram2}(d) is proportional to $(1-x)^3$, and the Raman peak areas match this cubic behavior.  The B$_3$O$_3$(OH)$_3$ concentration has a broad plateau in the range
$0.4 \le x \le 0.7$.  This plateau corresponds to the range where the monomers concentration drops in an approximately linear fashion [figure \ref{ram2}(e)].  Similar to the monomers, we also find the
B$_3$O$_3$(OH)$_3$ trimers in room temperature glasses.  However, the trimer concentration decreases much more abruptly to zero near $x = 0.36$.  This sudden decrease reflects a critical bond percolation point below which, with the exception of fluctuations, all boron atoms participate in the network glass.  The proximity of the points where trimers and monomers vanish ($x=0.36$ and $x=0.38$, respectively) suggests that they are both coupled by the percolation process.

Lines near 787 or 808 cm$^{-1}$ appear in all spectra with the exception of the composition endpoints $x=0$ and $x=1$. In this frequency range one can distinguish a narrow peak centered at 808 cm$^{-1}$ with a width of
$6 \pm 2$ cm$^{-1}$  and a broader peak that is $21 \pm 6$ cm$^{-1}$ wide.  The narrow component, centered at 808 cm$^{-1}$, corresponds to B$_3$O$_6$ boroxol rings and the composition dependency of the intensity is shown in figure \ref{ram2}(c).  This sharp line appears both in the vitreous spectra
and the liquid spectra with $x\le0.786$.
For low-$x$ glasses we find that the intensity of the 808 cm$^{-1}$ line is relatively small, and in general its intensity increases as $x$ approaches 0.5.  Similarly, in liquids the intensity increases as one approches $x=0.50$ from above.  However, the observed intensities are highly variable; these variations may reflect that the number of boroxol rings depend on the thermal history of the samples with deviations from thermal equilibrium.

\begin{figure}[ht]	           
\includegraphics[scale=1.0]{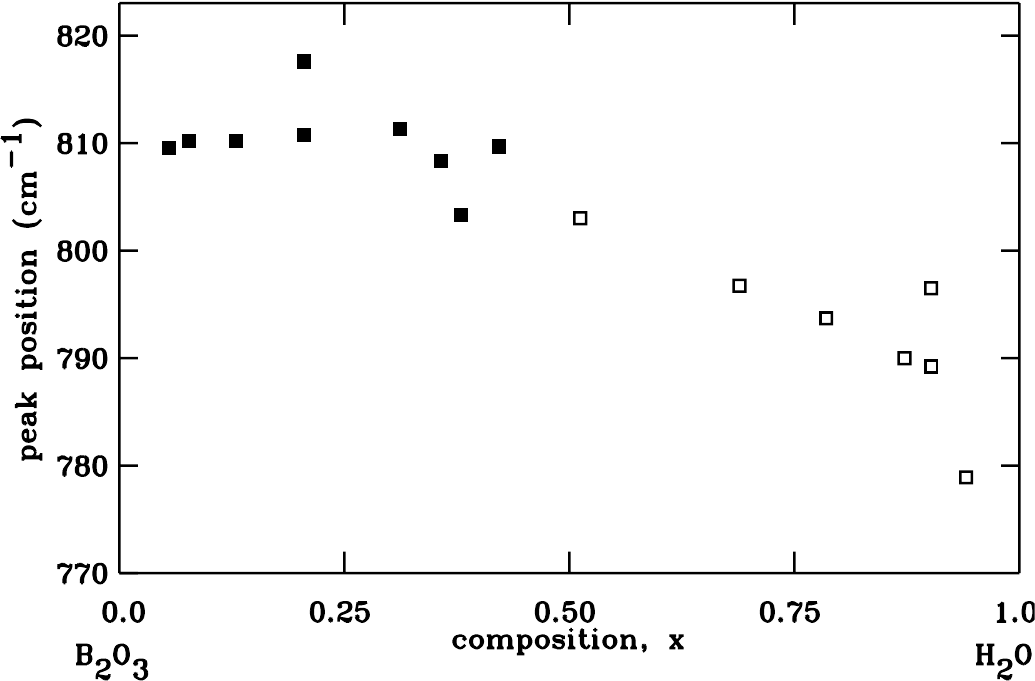}
\caption{\label{ram3}
Position of the broad component of the Raman peak centered near $\nu = 800$ cm$^{-1}$,
assigned to B$^{[4]}$--O stretching modes, as a function of composition.  Closed and open symbols indicate vitreous and liquid samples, respectively.
}
\end{figure}

The broad peak component in range between 787 and 808 cm$^{-1}$ is distinct from the narrow boroxol peak.  The intensity of the broad component is shown in figure \ref{ram2}(b), and the corresponding peak positions are shown in figure \ref{ram3}.  From the present data it is not clear whether this feature is a single line that gradually shifts its position for the liquid samples, or whether these are two lines, one of which vanishes as the other appears.  Raman frequencies in this range are related to oscillations involving tetrahedrally coordinated boron atoms (symmetric B$^{[4]}$--O stretching modes)  \cite{sch,jun,jan}.  These modes are most prominent in glasses with compositions in the range from $x \approx 0.25$ to 0.4, but they are present at all compositions with the exceptions of the water-free and the pure water samples.  The mean line position shifts towards lower frequencies with increasing $x$  (figure \ref{ram3}).  This shift of the Raman frequency occurs for the most part in the liquid state, and this decrease of the Raman resonance frequency may correspond to the switchover of the B$^{[4]}$--O bond environment from a boron--oxygen network to an aqueous surrounding as $x$ approaches one.

In addition  we observe a  line in the range
from 1650 to 1652 cm$^{-1}$.
The width of this line is in the range from 2 to 3 cm$^{-1}$, and it is thus narrower than the other features in the Raman spectrum.  Furthermore its peak height is comparatively low.
The composition dependence of the line is shown
in figure \ref{ram2}(a): the maximum of the intensity is located near $x=0.20$.  Based on this
composition dependence this line is associated with a vibration mode of isolated hydroxyl groups in the vitreous B$_2$O$_3$ network.

Spectra with $0.056 \le x \le 0.690$ exhibit varying amounts of fluorescent scattering.  A series of peaks in the range from 300 to 1400 cm$^{-1}$ is associated with this fluorescence, and a part of this sequence can be seen in figure \ref{ram1}.  This figure shows that these regularly spaced peaks are notably absent both in the pure glass as well as the water-rich liquids.  The width of these peaks increases gradually with increasing wavenumber.

\section{Discussion}	

One feature that emerges from both the x-ray scattering and Raman spectroscopy is the onset of polymerization at about $x=0.80$.  At this point the number of monomers reaches a maximum
[figure \ref{ram2}(e)], the number of trimers begins to deviate from the cubic behavior expected for a dilute solution [figure \ref{ram2}(d)], and the composition dependence of the position of the FSDP has a well-defined change of slope  [figure \ref{liquidus}(b)].

The point where polymerization is complete, i.e. where the whole sample becomes a single macromolecule, can be discerned from the Raman data:  at $x \approx 0.36$ the signal corresponding to the B$_3$O$_3$(OH)$_3$ trimers drops to zero in figure \ref{ram2}(d).  This point is well within the glass forming range.  Accordingly, water-rich glasses near the limit of the glass forming range do contain a considerable number of trimers as well as traces of B(OH)$_3$ monomers.  This appears reasonable since these water-rich glasses may coexist in metastable equilibrium with boric acid crystals consisting of sheets of B(OH)$_3$ molecules \cite{mcc}.  Bond percolation at $x \approx 0.36$ does not lead to distinct changes in the x-ray scattering patterns of the glasses in figure \ref{all}(b).  However, the appearance and proliferation of small molecular weight borate species is likely the origin of the increase in the excess specific heat and the switchover from strong to fragile behavior for glasses with $x>0.3$ \cite{bru4}.

The distributions of different molecular species in figure \ref{ram2} are consistent with the average stoichiometry of the sample.  For the composition  $x=0.75$ one finds, in addition to the B(OH)$_3$ monomers with $x=3/4$, a considerable concentration of B$_3$O$_3$(OH)$_3$ trimers with $x=1/2$.  This requires necessarily the presence of molecular water, in agreement with the spectroscopic results.
Similarly, for the average composition $x=0.5$ one detects, in addition to the $x=1/2$ trimers, both monomers and the spectroscopic features of more highly condensed molecular structures such as B$_3$O$_6$ boroxol rings.

The density of (B$_2$O$_3$)$_{1-x}$(H$_2$O)$_x$ liquids and glasses have been reported \cite{mil1,kod}.  The slope of the density-composition curve changes near $x=0.25$, such that for $x > 0.25$ some local atomic configurations appear that fill space less efficiently.   An associated feature in this composition range is higher small-$q$ scattering for the $x=0.299$ and $x= 0.445$ glasses [figure \ref{all}(b)].  The associated density fluctuations may reflect a structural heterogeneity that is consistent with the change of slope seen in the density data.  The start of the rise of the number of boroxol rings  [figure \ref{ram2}(c)] coincides with the decrease of the macroscopic density, which suggests these boroxol rings may be one cause of less efficient space filling.  Four-fold coordinated boron atoms are associated with denser packing in e.g.\ alkali borates, a phenomenon that is also referred to as the borate anomaly.  The maximum number of four-fold coordinated boron atoms is found here near $x=0.35$.  This maximum is relatively broad, and the overall signal is relatively weak, which limits the impact on the density.

\section{Conclusions}	
The structure and properties of amorphous and liquid phases, spanning the range from the B$_2$O$_3$ network glass to liquid water have been explored using x-ray scattering and Raman spectroscopy.  Raman spectroscopy provides clear information for the progressive polymerization of the low-weight molecular species with decreasing $x$. The  onset of polymerization can be identified at around $x=0.80$.  At this composition, the number of B(OH)$_3$ monomers assumes a maximum, whereas the number of trimers begins to deviate from the dilute solute $(1-x)^3$ behavior.  In the x-ray scattering data this point is marked by a change in the behavior of the FSDP; for lower $x$ the position of the FSDP remains approximately constant, while it shifts linearly to higher $q$ for higher $x$.  The critical point for bond percolation is clearly marked by the sudden disappearance of B$_3$O$_3$(OH)$_3$ trimers at $x=0.36$.  This point is located well within the glass forming range, and therefore water-rich glasses may contain trimers and even B(OH)$_3$ monomers.  Molecular water could only be detected in melts with $x\ge 0.61$.

\section{Acknowledgments}	
Support through a Discovery Grant from the Natural Science and Engineering Research Council of Canada is gratefully acknowledged.
\bibliography{bruening01}

\end{document}